# Progress in Brain Computer Interfaces: Challenges and Trends

Simanto Saha[1], Khondaker A. Mamun[2], Khawza Ahmed[3], Raqibul Mostafa[3], Ganesh R. Naik[4], Ahsan Khandoker[5,6], Sam Darvishi[1], and Mathias Baumert[1]

[1]The School of Electrical and Electronic Engineering, the University of Adelaide, Adelaide, Australia
[2]The Department of Computer Science and Engineering, United International University, Dhaka, Bangladesh
[3]The Department of Electrical and Electronic Engineering, United International University, Dhaka, Bangladesh
[4]The Marcs Institute for Brain, Behaviour and Development, Western Sydney University, Sydney, Australia
[5]The Department of Biomedical Engineering, Khalifa University of Science, Technology and Research, Abu Dhabi, United Arab Emirates
[6]The Department of Electrical and Electronic Engineering, the University of Melbourne, Melbourne, Australia

Corresponding author: Mathias Baumert (e-mail: mathias.baumert@adelaide.edu.au).

**ABSTRACT** Brain computer interfaces (BCI) provide a direct communication link between the brain and a computer or other external devices. They offer an extended degree of freedom either by strengthening or by substituting human peripheral working capacity and have potential applications in various fields such as rehabilitation, affective computing, robotics, gaming and artificial intelligence. Significant research efforts on a global scale have delivered common platforms for technology standardization and help tackle highly complex and nonlinear brain dynamics and related feature extraction and classification challenges. Psycho-neurophysiological phenomena and their impact on brain signals impose another challenge for BCI researchers to transform the technology from laboratory experiments to plug-and-play daily life. This review summarizes progress in BCI field and highlights critical challenges.

## I. INTRODUCTION

The brain computer interface (BCI) is a direct and sometimes a bidirectional communication tie-up between the brain and a computer or an external device, which involves no muscular stimulation, and has shown promise for rehabilitating subjects with motor impairments as well as for augmenting human working capacity either physically or cognitively. BCI was originally envisioned as a potential technology for augmenting/replacing existing neural rehabilitations or serving assistive devices controlled directly by the brain [1], [2]. The first systematic attempt to implement an electroencephalogram (EEG)-based BCI was made by J. J. Vidal in 1973, who recorded the evoked electrical activity of the cerebral cortex from the intact skull using EEG [1]. Another early endeavor to establish direct communication between the brain and the computer of people with severe motor impairments had utilized P300, an event related potential of the brain [3]. As an alternative to conventional therapeutic rehabilitation for motor impairments, BCI technology seeks to artificially augment or re-excite synaptic plasticity in affected neural circuits. By exploiting undamaged cognitive and emotional functions, BCI aims at re-establishing the link between the brain and an impaired peripheral site [4]. However, the research application of BCI technology has evolved over the years, including brain fingerprinting for lie detection [5], detecting drowsiness for improving human working performances [6], controlling virtual reality [7] and video games [8], and driving humanoid robots [9], [10], [11]. Figure 1 shows the increasing number of publications on BCI in various application fields over the years since its first appearance in a seminal article in 1979 [1].

According to the BNCI (Brain/Neural Computer Interaction) Horizon 2020 project, an initiative by the European Commission for coordinating BCI research community, six major application themes, i.e., restore (e.g., unlocking the completely locked-in), replace (e.g., BCI-controlled neuroprosthesis), enhance (e.g., enhanced user experience in computer games), supplement (e.g., augmented reality glasses), improve (e.g., upper limb rehabilitation after stroke) and research tool (e.g., decoding brain activity with real-time feedback) have been outlined as feasible and promising fields of BCI application [12].

### A. CHARACTERIZATION OF BCI SYSTEMS

BCI systems can be categorized by the way they use the brain: Passive BCI decode unintentional affective/cognitive states of the brain [11], while active BCI directly involve the user's voluntary intention-induced brain activity. Reactive BCI use brain waves generated as reactions to external stimuli. Detecting driver's drowsiness to prevent road accidents is an example of passive BCI [13]. BCI systems driven by user's intentional motor imagery [14] and visually evoked P300 produced by external stimulation [3], [5] can be considered as active BCI and reactive BCI, respectively.

The modality of signal acquisition has been used to divide systems into invasive and noninvasive BCI [15], [16]. Noninvasive BCI exploiting EEG are most common, although more recently, functional near infrared spectroscopy (fNIRS) [17], magnetoencephalogram (MEG) [18] and functional transcranial Doppler ultrasonography [19] have been exploited. In contrast, invasive intracortical electrodes [20] and electrocorticogram (ECoG) [21] have been used, providing a superior signal-to-noise ratio and better localization of brain activity.

Recent advancements allow both the decoding of neural activities and the delivery of external signals into targeted brain areas to induce plasticity [22]. While most of the BCI systems translate brain signals as computer command, some systems utilize external stimulation modalities such as transcranial magnetic stimulation [23], [24] and transcranial direct current stimulation [25] to stimulate specific brain areas. The bidirectional framework of BCI comprises of either one brain with feedback modality or two brains. Transcranial direct current stimulation directed by motor imagery-related EEG signals alters the connectivity in the sensorimotor networks in healthy individuals [26]. Another possible application of bidirectional BCI framework is direct brain-to-brain communication [23], [24]. Figure 2 illustrates a typical bidirectional BCI framework. Moreover, some BCI applications require auxiliary modalities, i.e., proprioceptive feedback and functional electrical stimulation driven by brain signals as feedback for augmenting or regaining peripheral motor actions [27], [28].

### B. FACTORS INFLUENCING BCI PERFORMANCE
For medical applications of BCI, three criteria are essential: (1) a comfortable and convenient signals acquisition device, (2) system validation and dissemination, and (3) reliability and potentiality of BCIs [29]. For the restoration of mobility in patients with motor impairments, invasive intracortical recordings show better BCI performance [30] than noninvasive methods such as EEG [31]. Invasive modalities are also suitable for locked-in patients, because the benefits (significantly improved quality-of-life) outweigh the risks associated with implantation [32]. A pilot study with a 1-year follow-up found no adverse effects pertaining to surgery or tissue reaction [33]. Invasive BCI should generally not be considered for neurologically intact subjects.

Many factors influence BCI performance; understanding the underlying neuronal mechanisms of cortical-subcortical networks is of crucial importance. For example, motor imagery-induced signals are recorded from premotor and motor areas, because premotor cortex, primary motor cortex and supplementary motor area along with basal ganglia and thalamus of the subcortical areas are the mostly activated areas during motor imagery [34].

Some issues can significantly impede BCI performance. Maintaining an acceptable signal-to-noise ratio in noninvasive long-term recordings is critical. Event-induced brain waves or oscillations are dynamic while being affected by unstable resting state networks (RSNs) [35]. Time-variant psychophysiological [36], [37], [38], neuroanatomical [39] factors and user's fundamental traits [40] cause unreliable estimates of RSNs, causing short and long-term signal variation within and across individuals. Due to these intrinsic signal variations, BCI systems require subject-specific training, in which subjects attend a calibration session that is tedious and often frustrating for the subject. To eliminate subject-specific training, the concept of inter-subject associativity, demonstrated in previous works in case of natural vision [41] and natural music listening [42], could be exploited toward inter-subject operable BCI. A recent study suggests that inter-subject operable BCI might become feasible with associate subjects [43]. Inter-subject BCI holds most promise for healthy subjects in applications like gaming, drowsiness and lie detection, because rehabilitative BCI must consider the characteristics and severity of individual impairment [44]. Another way to reduce the effects of session-to-session and subject-to-subject variabilities is transfer learning, in which algorithms are designed to test on datasets that are either not or minimally used to train the algorithms [45].

## II. CHALLENGES
Table 1 and Figure 3 highlight the fundamental BCI challenges in psycho-neuro-physiological and technological domains.

### A. PSYCHOPHYSIOLOGICAL AND NEUROLOGICAL CHALLENGES
Underlying emotional and mental processes, neurophysiology associated with cognition and neurological factors, i.e., functions, anatomy, play crucial roles in BCI performance and give rise to significant intra- and inter-individual variability. Psychological factors including attention, memory load, fatigue and competing cognitive processes [36], [46], [47] as well as users' basic characteristics such as lifestyle, gender and age, [39] influence instantaneous brain dynamics. For example, individuals with lower empathy participate less emotionally in a P300-BCI paradigm and can produce higher amplitudes of P300 waves than subjects with greater empathetic involvement [48]. Motivation is also related to P300-BCI performances and, such BCI is only applicable for people with intact visual perceptions [49].

Besides psychological traits, resting state physiological parameters, for example, frequency domain features extracted from resting state heart rate variability analysis are associated with BCI performances [50]. In addition, the baselines of RSNs are dynamic and modify any cortical signature instantaneously [35]. Age alters RSNs and associated cognitive responses [51]. Adapting with such time-variant RSNs is more demanding when the effects of RSNs are prominent over event-related cortical responses [52]. Moreover, the inherent complexity and diversity in the formation of a human brain [53] that influence the functional neural networks [54], construct highly volatile neuronal connectivity over time and across subjects [55]. An efficient BCI system must be robust to such inherent physiological fluctuations over time courses and, BCI without subject-specific calibration can promote more generalized frameworks for comforting healthy users.

Experiments correlating BCI performance with neuroanatomical, neurophysiological and psychological parameters have provided fascinating results: gray matter volume in sensorimotor cortical areas is associated with BCI success [39]. Sensorimotor rhythm-based BCI has implicated that physiological predictors such as spectral entropy and power spectral density, derived from resting state EEG recordings are correlated with BCI performance [37], [38]. Psychological predictors such as attention and motivation, are also associated with sensorimotor rhythm-based BCI performance [56]. Complementing EEG-based and questionnaire-based psychological predictors, corticospinal excitability could be used as another reliable marker for BCI performance [57]. Taking head anatomy into consideration augments BCI performance [58].

Around 15-30% of individuals are inherently not able to produce brain signals robust enough to operate a BCI [59]. Considering neurophysiological phenomena may reduce BCI illiteracy. An adaptive machine learning approach incorporating neurophysiological and psychological traits has been proposed to reduce BCI illiteracy [60].

Other case-specific investigations on neuro-psycho-physiological parameters contributing to BCI performance are essential. For the rehabilitation of stroke survivors, affected neural circuits, i.e., lesions are to be identified carefully, because brain responses fluctuate according to the spatial location of the stroke lesion [44]. Although current neuroimaging methods are effective in capturing stroke lesion sites, a case-specific BCI design that incorporates residual brain function is required for rehabilitative interventions. Highly individualized design impedes wide dissemination of BCI-driven rehabilitation of neurological conditions.

### B. TECHNOLOGICAL CHALLENGES

Event related potential (ERP) [61], steady-state visual evoked potential (SSVEP) [62], auditory evoked potential (AEP) [63], steady-state somatosensory evoked potential (SSSEP) [64] and motor imagery (MI) [14], have been proposed to detect cognitive signatures although none of the signatures performs well for all BCI applications. For example, ERPs and SSVEPs are target-specific and generated by external stimulation; however, if ERPs depend on visual stimuli, they cannot be used for rehabilitating locked-in patients with impaired visual processing. In that case, auditory-based ERP could be used if auditory processing remains intact. On the other hand, MI seems too slow for action control, thus they are not suitable for controlling virtual reality environments or videogames [65]. Recently proposed hybrid BCIs which utilize more than one waves, i.e., SSVEP/ERP [66], [67] and SSVEP/MI [68], seem to offer more robust features.

Intrinsic neurophysiological instability of the brain dynamics poses critical challenges for making BCI systems efficient. The major components of a BCI system are the signal acquisition, signal processing and effector device [69]. Various neuroimaging techniques have already been in use to explore cortical activities as forms of either electrical or hemodynamic signatures [16] and none of the methods shows any advantageous outcome for a lucrative BCI design complying four important criteria, i.e., cost efficiency, portability, easy maintenance and less or no involvement of invasive surgery. Invasive signal acquisitions are yet to be tested for long-term reliability for assessing whether the placed sensors could maintain the signal quality over a long period inside the brain [70].

EEG [71] provides relatively poor spatial resolution due to noninvasive scalp recordings compared to fMRI, but finer temporal resolution [32], [72]. Employing high density EEG mapping to increase spatial resolution results in high computational cost and efforts to maintain a reasonable signal-to-noise ratio across all channels [73]. Since EEG captures only the electrical field generated associated cognitive processes, concomitant assessment of blood-oxygen level-dependent (BOLD) activity may improve BCI performance. BOLD activity is typically captured with fMRI [74], which is not feasible for most BCI applications, due to unmanageable size and cost of the device. fNIRS provides a safe, noninvasive, relatively inexpensive and portable neuroimaging alternative for recording BOLD activity [75], [17]. Integrating fNIRS with EEG can significantly enhance classification performances regardless of low information transfer rate caused by inherent delays in hemodynamics [76]. A recent study has suggested that fNIRS is unable to adequately offer acceptable performances on its own, but can be combined with EEG to boost the performances [77]. Probing cortical sources is another important issue for scalp-based sensors such as EEG, as they could be located subcortically. Reconstructing task-induced sources while resolving the so-called inverse problem imposes a significant challenge. A two-equivalent-dipole model was applied on EEG data to discern the anatomical nature of the MI induced sources and to aid the classification performances [78]. The neuronal signals attenuate through some layers of tissues having complex geometry and diverse electrical properties; however, the magnetic permeability in the cerebrospinal fluid, skull, and skin, is consistent [79]. Thus, MEG is more likely to capture signal with lesser distortion than EEG. Even MEG provides better spatiotemporal resolution as compared to EEG, although the use of this technology is limited by cost and portability [80].

The design of classifiers for BCI comprises two issues [81], [82], [83]. First, the dimensionality of the features set used for estimating the model parameters should be chosen for optimal performance based on the nature of the classifier. Second, the trade-off between bias and variance has to be considered and may involve regularizing the parameter estimation.
Covariate shift of class specific feature sets during training and testing a classifier model, following unlike probability distributions although their conditional distributions remain unchanged [84], is an important issue requiring the application of adaptive methods for compensating feature space transitions. The unsupervised subspace learning method enables session-to-session and subject-to-subject information transfers, augmenting BCI performance [85], [86], [44]. As a supervised method, common spatial pattern has been extensively used in EEG-based online and offline BCI settings [87]. However, a common problem with such a data-driven technique is over-fitting of the model parameters based on training sets, causing unreliable prediction on the test data [88].

## III. NEURAL PLASTICITY, SENSORS, SIGNAL PROCESSING, MODELING AND APPLICATIONS
Exploiting neural plasticity, designing hi-fidelity and customized neural sensors, applying advanced signal processing, and machine learning techniques are the key aspects of BCI design. Table 2 summarizes a set of applications of BCI technology.

### A. SYNAPTIC PLASTICITY AND COGNITIVE REHABILITATION
The time-variant behavior of synapses within complex neural networks underpins the plastic characteristics of the brain and was first illustrated by Donald O. Hebb in 1949 [89]. Closed-loop BCI with neurofeedback is assumed to contribute to the reorganization of cortical-subcortical neural networks and assist subjects in self-regulating specific brain rhythms; notwithstanding, the underlying mechanisms that alter neural substrates are still not well-understood [90]. For example, BCI-based covert visuomotor training modulates associated neural substrates [91], where the effects of modulated neural substrates are observed while performing that particular movement-related task. Substantial changes in overt movement-related task following BCI-driven training induced learning suggest a critical role of BCI in enhanced motor learning for proficiently controlling neuroprosthetics [92]. BCI may augment training-induced plasticity during therapeutic motor rehabilitation and, thus, re-excite corresponding neural substrates to regain control by means of neuroprosthetics or upper limb functions [93].

The extent of BCI-induced plasticity entails several factors, including (1) the selection of the signal acquisition modality, which plays important role in diagnosing neural states, (2) the design of feedback modality that has explicit association with the neural signal classification performance, (3) the consideration of application-specific feedback delays, and (4) the utilization of a suitable feedback modality [94]. Neural ensemble recordings using signal acquisition modalities such as EEG, MEG, fNIRS and fMRI has become dominant over single unit recordings. Behavioral activities are likely to be distributed across three-dimensional cortical-subcortical networks and that cannot be captured within single unit recordings [95].

Rehabilitative BCI can be designed either by attaching neural prostheses to the impaired body parts or by re-stimulating the damaged synaptic networks; in any of the cases, the idea is to exploit and promote neural plasticity [96]. In stroke patients with paretic muscles without residual finger movement, increased electromyogram activity post rehabilitation by BCI-driven orthoses exhibits increased neuromuscular coherence that is essential for restoring movement control [97]. Explicit application of functional electrical stimulation regulated by EEG-based movement-related signatures further suggests a role of BCI in rehabilitation [98]. Increased electromyogram activity in paretic muscles is indicative of plasticity induced by electrical stimulation [28]. For BCI-based rehabilitation in a real-life environment, differentiating between task-induced activities and resting state activities is a key factor for controlling the prosthesis or stimulation modality [99].

Externally stimulating the affected brain areas by electric or magnetic fields holds promise for stroke rehabilitation. A recent study demonstrated the induction of neural plasticity in white matter and cortical functions in chronic stroke patients by motor imagery-based BCI and transcranial direct current stimulation applied to targeted brain areas [100]. Magnetic stimulation of brain areas driven by BCI increases cortical activation in stroke patients [101]. Plasticity alteration post-rehabilitation varies across subjects and, thus, an individual-specific training session is necessary [102]. The usefulness of BCI-based motor rehabilitation is limited for locked-in patients because they are unable to interact with the system [103]. Other examples of BCI-driven rehabilitations include optimizing the parameters for deep brain stimulation applied into the subthalamic nucleus in patients with Parkinson's disease [104] and treating major depressive disorder patients by BCI-driven transcranial magnetic stimulation [105].

Either by providing direct control of assistive technologies or by neural rehabilitation, BCI can help patients who may suffer from amyotrophic lateral sclerosis, cerebral palsy, brainstem stroke, spinal cord injuries, muscular dystrophies, or chronic peripheral neuropathies [106]. Providing auxiliary degrees of freedom improves quality of life significantly for people with disabilities. Brain signals can be translated to drive wheelchairs [107], [108]. Integration of BCI with a vision-guided autonomous system was shown to effectively perform the grasping task using a prosthetic arm in a Tetraplegia patient [109]. An implanted microelectrode array has been proposed to operate a three-dimensional neuroprosthetic device [110].

### B. SIGNAL ACQUISITION, SIGNAL PROCESSING AND MODELING
Control signals exploiting either cognitive hemodynamics or electrical activities can be used to operate a BCI efficiently. A significant number of studies are now involved in combining multimodal signal acquisition modalities to augment current BCI systems. For example, simultaneous EEG and fMRI yield complementary features by exploiting good temporal resolution of EEG and good spatial resolution of fMRI [111]. Enhanced multiclass sensorimotor tasks classification performance using hybrid EEG and fNIRS signals implicates the importance of features extracted from both hemodynamic and electrical activities [112]. MEG is another potential tool to combine with EEG, as it captures radially/tangentially dipole sources in cortical-subcortical networks and adds complementary information to EEG signals [113]. Hence, different signal acquisition modalities could be combined together to improve the BCI efficiency.

The combination of signal processing and machine learning approaches plays critical role in translating any brain signal to a command for a computer or other external devices. Here, we only highlight signal processing and machine learning techniques for processing EEG, ECoG and fNIRS signals. Representing signals in the time-frequency-space is necessary to obtain physiological correlates of BCI outcomes [81]. Fourier transform (FT) and autoregressive models are examples of time domain representations while short time FT and wavelet transform allow for time-frequency representations of brain signals [114], [81]. In case of spatial filtering, the most popular filtering approaches are common spatial pattern, independent component analysis and the Laplacian filter. A diverse range of inverse models allow to discern the actual sources projected on three-dimensional cortical-subcortical networks. Extracted features can be translated using various linear and nonlinear classification algorithms. Examples of linear and nonlinear classifier models are linear discriminant analysis and nonlinear kernel-based support vector machines [82], [83].

Since the first publication in 2000, common spatial pattern is still one of the most popular methods to represent multichannel EEG signals by corresponding spatial contents [87]. As a data-driven method, it requires a significant number of training samples to model the filtering parameters. In case of small training trials, regularizing the covariance estimation works better than the traditional algorithm [115]. Other modifications in spatial filtering include projecting EEG by using sparse representation and filter bank spectral division of raw signals [116]. Generally, spatial filtering is applicable in subject-specific BCI development although a recent study has proposed estimating the filter coefficients from a subject and applied that filter to another subject, which contributed no training sample [43]. On the other hand, independent component analysis is a blind source separation method requiring no training. The estimation of independent components are based on statistical properties of the signals [117]. However, modeling the actual cortical sources as dipoles in the complex brain anatomy from the scalp EEG recordings seeks to solve the so called inverse problem [118], [119]. More recent source localization methods such as wavelet-based maximum entropy on the mean represent EEG/MEG signals as relevant time-frequency contents and finally transform them into spatial representations [120].

### C. NEUROSENSORS: THE-STATE-OF-THE-ART
Deeper regions of the brain, e.g., subcortical and cerebellar regions, contribute to various neuronal activities [121], [122]. Interpreting the genesis of cortical sources from cellular to scalp levels and RSNs spanned throughout the three-dimensional brain space can guide BCI development [123]. Sensors with customized design are developed to advance brain signal acquisition modalities. Neurosensors can be constructed in different forms like electrical, optical, chemical and biological [124]. Dry EEG electrodes are convenient, but provide inferior signal-to-noise ratio compared to conventional wet electrodes, which use conductive gel and require proper skin preparation minimizing the skin-electrode impedance [125]. However, a study on dry electrodes-based BCI suggested that dry electrode could be used to collect good quality signals by designing the circuits carefully [126]. While utilizing the advantages of both dry and wet electrodes, quasi-dry electrodes exploiting the mechanical properties of polymer can capture signals as comparable to commercial Ag/AgCl electrodes [127]. To increase the spatial resolution of EEG, Petrov et al. have proposed an ultra-dense sensor array of 700-800 electrodes [128]. The signal-to-noise ratio was twice as high as for high-density EEG that has up to 256 gold-coated electrodes. An auricle electrode with stretchable connector was proposed that not only can increase portability but also can offer a comfortable alternative for long term recordings [129]. The electrode is flexible with the alterations of electrical and mechanical properties of skin, thus it would be a comfortable option for long term recordings.

Invasive signal acquisition sensors must be biocompatible. A novel organic electrochemical transistor-based sensor enables to collect neural signals directly from the brain surface [130]. This sensor is biocompatible and mechanically flexible, and the transistor-based design amplifies captured signals locally, thus providing much better signal-to-noise ratio than conventional ECoG. To enhance the signal quality, carbon nanotube coating can decrease the electrode impedance and, thus, increase the charge transfer [131]. Another invasive biocompatible sensor, designed for recording previously inaccessible spectra of large neuron populations, includes data transmission for use in natural environments [132]. Oxley et al. have proposed stent-electrode array (stentrode) that involves minimal invasiveness [133]. Using computer-guided catheter angiography, the stentode can be placed within arteries or veins located inside the brain anatomy. Capturing high-fidelity cortical signals, this technology will significantly reduce the risk factors of craniotomy. With the outstanding progression of nanotechnology, nanowire Field Effect Transistor and other p/n junction devices have potential for neuro-sensing modalities for intracellular recordings, even in the deep brain regions [134].

Besides large scale recording modalities like EEG and MEG, very small scale recordings of neuronal activities is crucial for understanding brain circuits' functions and intra- and inter-neuron interactions. Representation of any cognitive task as functions of both small scale and large scale neuronal interactions is crucial for the advancement of neuroengineering and BCI. In this regard, a high-density neurosensor array made from silicon probes combined with optogenetics enables single unit recordings [135]. Yang et al. proposed a novel multi-plane two-photon microscope that can be used to capture multi-layer neuronal structure and mechanism with cellular resolution [136]. Other potential imaging methods for investigating cell signaling include calcium imaging [137] and advanced microscope with chronically implanted lenses [138]. Designer receptor exclusively activated by designer drugs, provides a chemogenetic tool to understand cell-signaling including electrical activities in molecularly clustered cell groups [139], [140]. A new ultrasonic-based wireless system, called neural dust, enables the recording of electromyogram and electroneurogram on the milimeter scale [141].

### D. MISCELLANEOUS APPLICATIONS
The applications of BCI are extremely diverse, specifically as an augmentative tool for healthy individuals. An EEG-based BCI-driven controller of mobile robot has demonstrated the possibility of this technology in robotics industry [142]. BCI can also be used for controlling humanoid robots remotely using EEG [9], [10], suitable in hazardous environments, for example by sending a robot in a coal mine for executing a task that is potentially unsafe for a human. In space, BCI can be used to monitor astronauts' working capacity [143] and to drive an exoskeleton [144]. In the absence of gravity, working becomes tedious and inconvenient. Furthermore, astronauts' working time is precious. BCI-driven systems could be practical for improving astronauts' functionality, efficiency and safety [145]. Other applications for healthy users consist of controlling virtual reality and video games [67], brain fingerprinting [5], [146], mood assessment [147], and brain painting [148].

More recently, brain-to-brain interface (BBI), involving decoding sender's cognitive intentions, translate them into commands for stimulating receiver's brain have been explored. In 2013, researchers implemented a direct BBI system in which one rat was able to share sensorimotor information to another rat [149]. Intracortical microstimulation was used to stimulate the receiver's target brain areas. An early attempt to develop sensorimotor rhythm-based BBI between two human subjects used noninvasive EEG and transcranial magnetic stimulation has been proposed by Rao et al. [23]. Other total noninvasive BBI experiments have proposed sharing pseudo-random binary streams encoding words between human subjects [24] and playing collaborative games [150].

## IV. ETHICAL CONCERNS AND SOCIOECONOMIC CONTEXTS

Irrespective of the scientific breakthroughs in BCI field, there are key factors pertaining to safety, ethics, privacy protection and data confidentiality, community acceptance and socioeconomic aspects that should be considered with adequate precautions to maximize users' benefits and social impacts [151], [152], [153], [154].

Physical and mental safety of BCI users is important. Invasive procedures such as deep brain stimulation may cause postoperative psychological and neurological side effects [155], [156], [157]. Additionally, bleeding and infections are infrequent but do occur and may require removal or further maintenance of the implanted electrodes. Guidelines are required to safely advance neurotechnologies [158], because BCI devices can alter behavior and, thus introduce potential threats to one's emotions, personality and memories; more generally one's mind. For human brain-to-brain interface applications [23], [150], one may define an upper bound for research depths keeping in mind the necessity of ethical utilization of this technology. Because both sender and receiver play complicated roles, more specifically, sender's intentional manipulative control over neural signals might alter the anticipated outcome. Altering human cognitive and possibly moral capacity raise a serious ethical question and it is not predictable if the cognitive changes reversible and efficacious [159].

A user's expectations of achieving extended or auxiliary degree of freedom may not be fulfilled, and even the unfamiliar risk factors can diminish the accomplished advantage of using BCI [160], [161]. Creating broad awareness of BCI technology and its pros and cons would educate people, who fear unnecessary technological dependency [162].

It is critical to introduce a suitable act for lawful utilization of BCI and preservation of privacy and confidentiality of stored data. Necessary precursor initiatives should propose application-specific BCI frameworks, which can restrict unauthorized access to stored data or the system [163]. For example, illicit access to a wireless BCI-driven limb and manipulative reprogramming of a computer-guided neuro stimulation have demonstrated the importance of establishing resilient safeguards to BCI use [164]. Without evaluating socioeconomic, ethical and policy issues, the commercialization of BCI would hinder BCI progress [165].

By creating a common networking platform for BCI researchers worldwide, the immediate proposition of a comprehensive list of universal guidelines is key to sustainable advancements of the field [166]. Various alliance-based projects are running as common platforms for advancing the knowledge of neuroscience [167]. The European Union along with its partner universities have initiated The Human Brain Project (HBP). In addition, the Brain Initiative has been announced by the White House. A comprehensive set of social and ethical guidelines have been described by Rose [168]. Figure 4 highlights the present states and predicted future achievements of BCI technology.

## V. CONCLUSION

Numerous groundbreaking advances in neurosensors and computational tools herald great promise for more sophisticated and user friendly BCI systems requiring no or little maintenance. In addition to hi-fidelity signal acquisition, significant progress in signal processing and machine learning tools, their complementary roles, and high computation power and increased mobility of computers have significantly contributed in the emergence of BCI technologies. The future of BCI technology will rely greatly on addressing the following key aspects:

- Elucidating the underlying psychophysiological and neurological factors that potentially influence BCI performance.
- Designing sensors with greater signal resolution, while considering portability, easy maintenance, minimal or no invasiveness and affordability.
- Estimating covariate shift for subject-specific feature spaces (i.e., training and testing feature sets) and modeling subject-to-subject information transfer for the proposition of more generalized BCI models with insignificant or no calibration requirement.
- Establishing broad consensus on ethical issues and beneficial socioeconomic application of this technology.

**ACKNOWLEDGMENT**
We would like to thank Prof. Moritz Grosse-Wentrup for providing his valuable feedback.

**CONFLICT OF INTEREST**
None declared.

# Figures and Tables

**FIGURE 1.**

The number of publications over the years: The statistics was based on a search on PubMed in which 'brain computer interface' was the search keyword. The publications those were listed until 17th May 2018 have been accounted only. A significant increase in the number of publications in this decade as compared to the last decade implicates the engagement of a greater community in this field and, thus the importance of BCI technology.

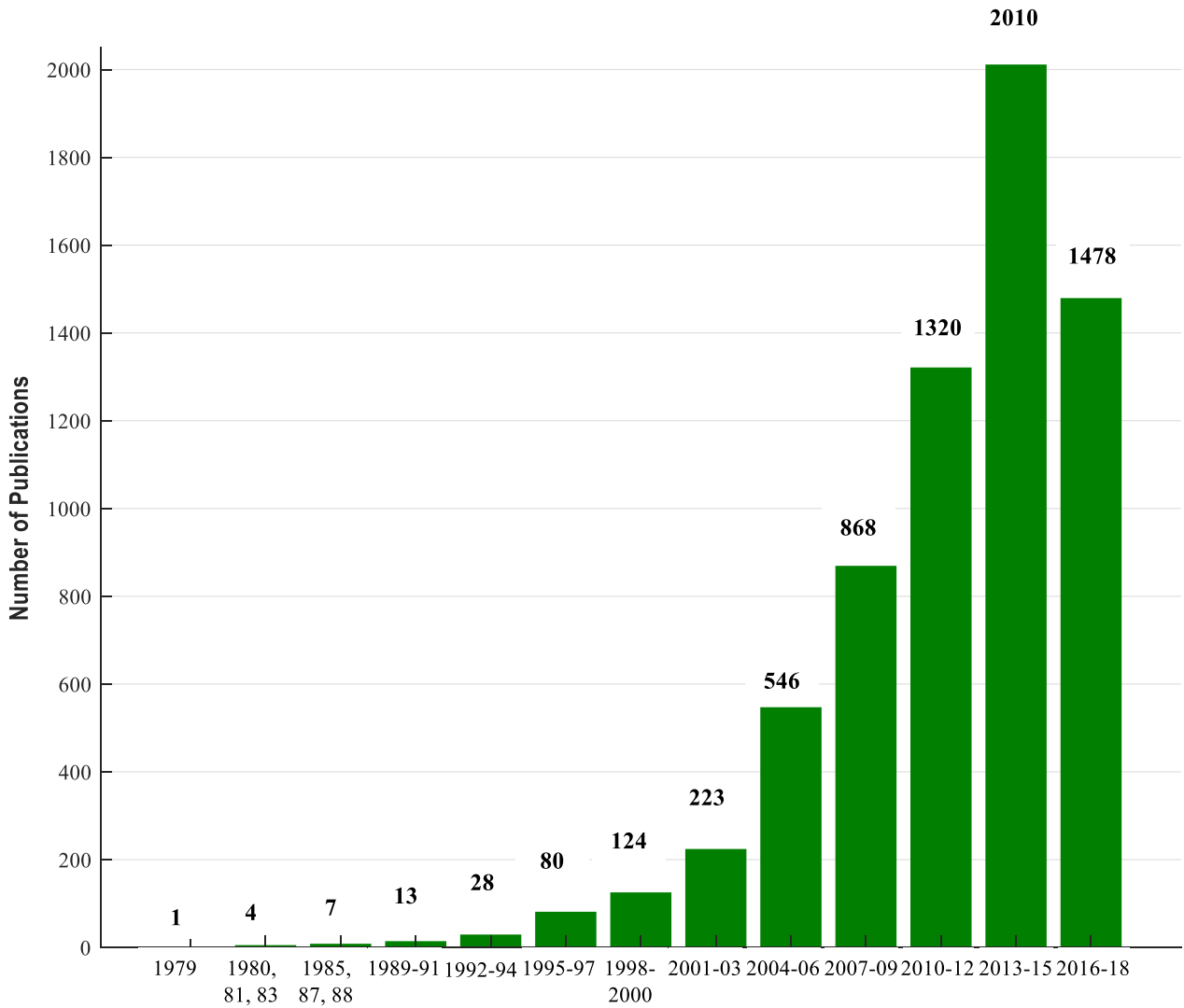

**FIGURE 2.**

Basic block diagram of communication between a brain and a computer: (a) a typical brain computer interface framework and its components, (b) a typical computer to brain interface framework and its components.

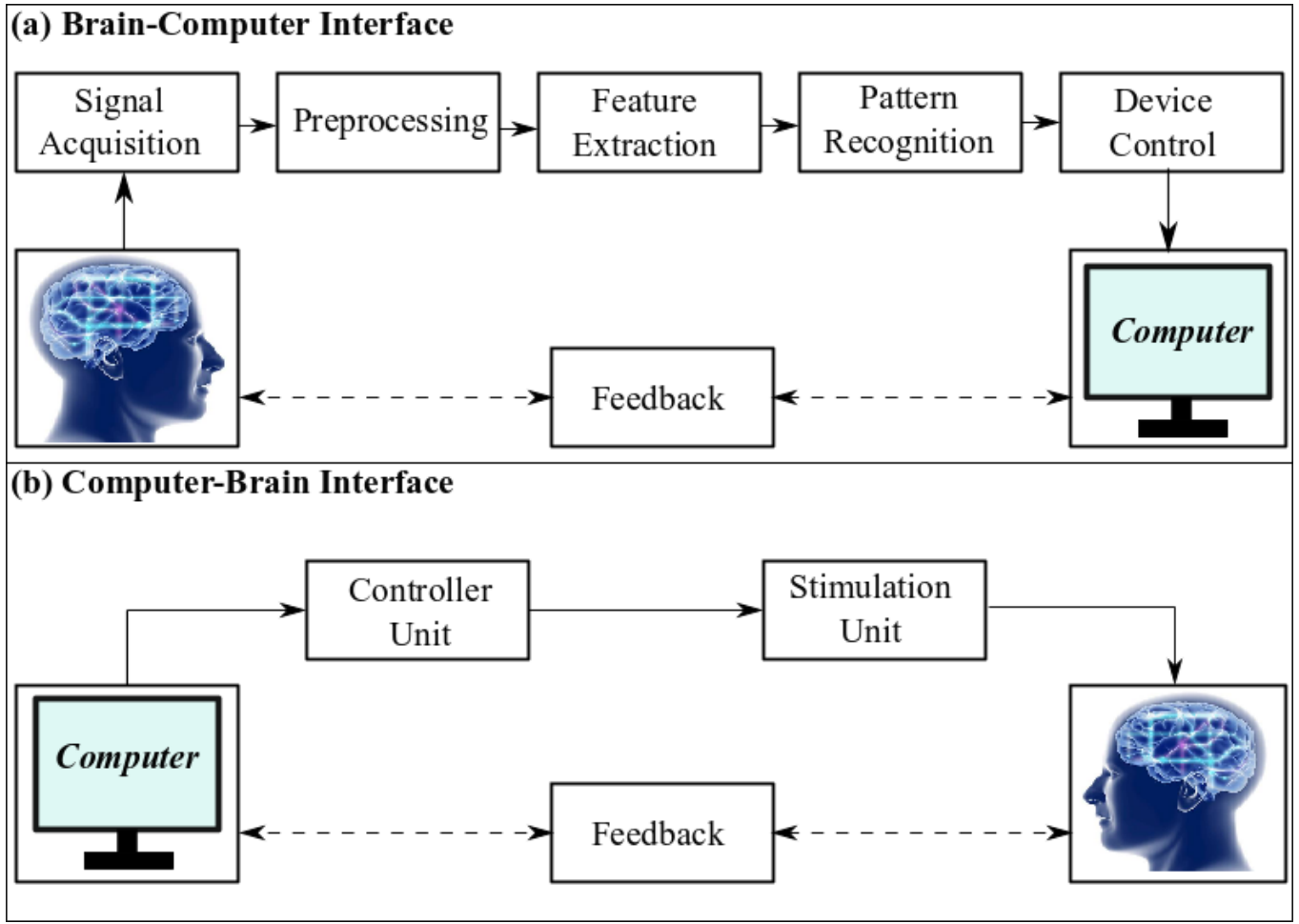

**FIGURE 3.**

The psychophysiological, technological and ethical challenges in brain computer interface development.

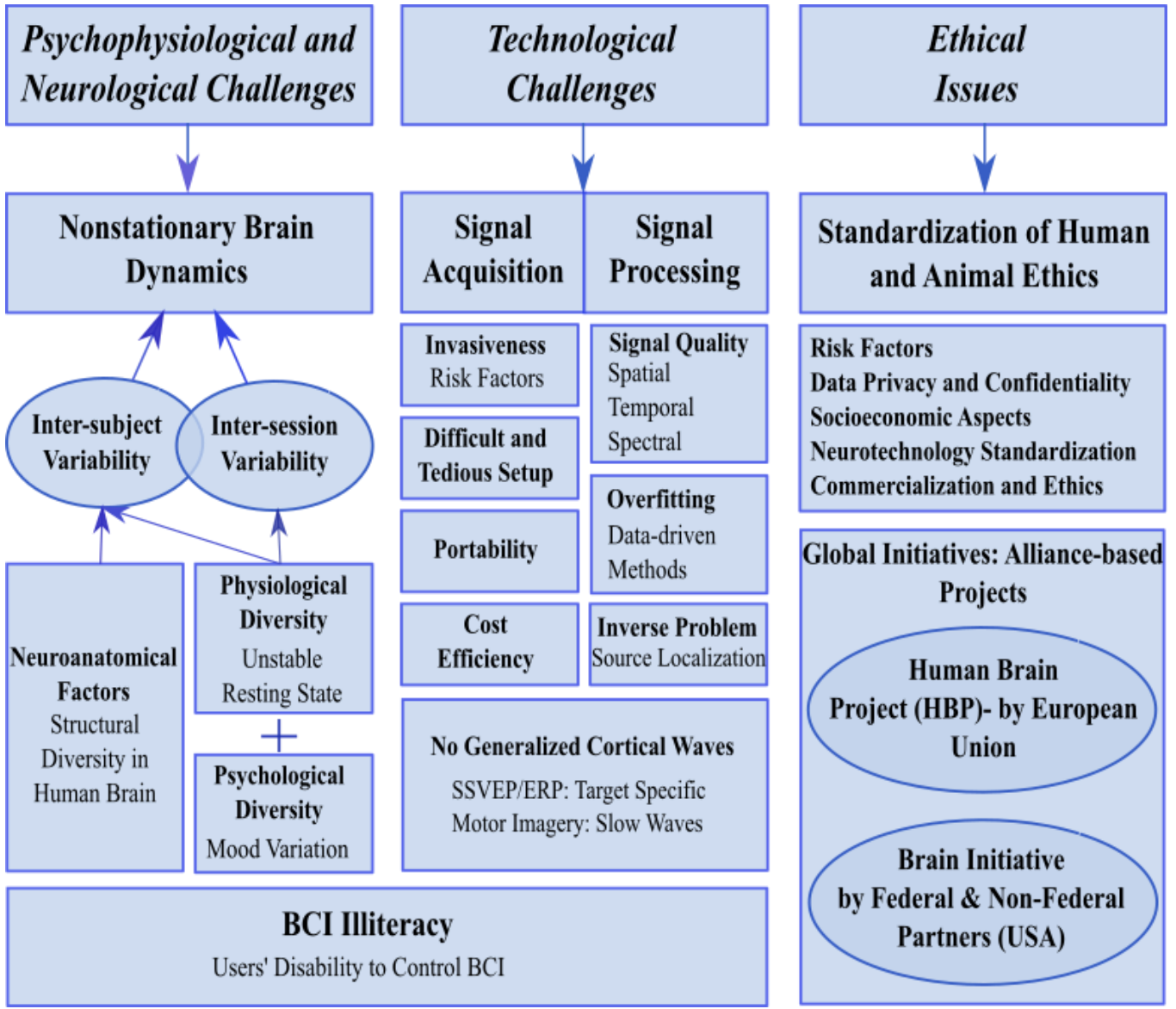

**FIGURE 4.**

The current states and future prediction of brain computer interface technologies: (a) predicted improvements in signal acquisition modalities and (b) predicted applications with extended degree of freedom. [Abbreviations: ECoG- electrocorticogram, EEG- electroencephalogram, fNIRS- functional near infrared spectroscopy; adopted from BNCI 2020 Roadmap].

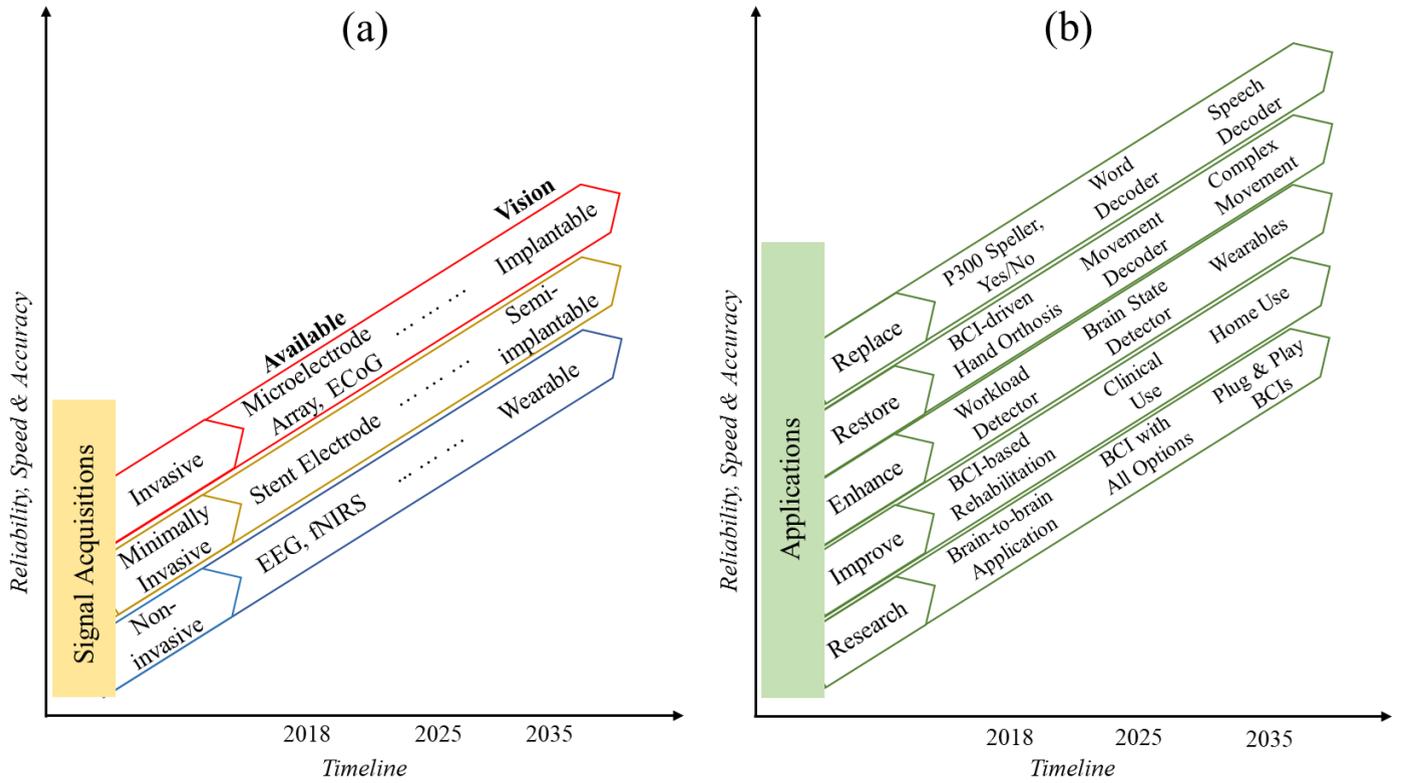

**Table 1**

SUMMARY OF BCI CHALLENGES IN PSYCHOPHYSIOLOGICAL, NEUROLOGICAL AND TECHNOLOGICAL DOMAINS

| *Domain* | *Challenges* | *Remarks* |
|---|---|---|
| **Psychological, Physiological and Neurological** | Cognitive variability over time and across individuals | Attention, memory load, fatigue and spontaneous cognitive processes continuously alter over time and across individuals, while introducing diversity in the event induced cortical activities [36], [46], [47]. |
| | Non-unique basic characteristics | Users' demographics such as gender and age, and way of living influence the nonstationary brain dynamics [40]. |
| | Resting state network (RSN) | RSN characterizes the underlying cognitive processes of brain and influence the event related activities. The baseline of RSN is unstable over short and long time courses and non-unique [35]. |
| | Neuroanatomical structure | Individuals possess diverse anatomical structure of the brain that is correlated with BCI performances [39], [58]. |
| | Spatially locating region of interest | Example: identifying lesion location for stroke survivors [45]. |
| **Technological** | Signal acquisition | None of the signal acquisition modalities is able to provide high resolution signals in spatial, spectral and temporal domains while maintaining the most important BCI features, i.e., efficiency, mobility, ease of use and safety, simultaneously [16]. |
| | Signal preprocessing, feature extraction and classification | While capturing brain signals from the scalp by noninvasive means, dealing with inverse problem is critical. Furthermore, non-brain artifacts degrade signal quality. No generalized signal processing, feature extraction and pattern recognition methods have been found yet [81], [82], [83]. |
| | Brain waves | Event related potential [63], steady-state visual evoked potential [64], auditory evoked potential [65] and steady-state somatosensory evoked potentials [66], are target specific and motor imagery [14] is a slow wave and, thus they not suitable for virtual reality and gaming applications [67]. |
| | BCI illiteracy | About 15% to 30% of BCI users are unable to control the systems due to BCI Illiteracy [59]. |

**Table 2**

LIST OF POTENTIAL APPLICATION FIELDS FOR BCI TECHNOLOGY

| *BCI application fields* | *Examples* |
|---|---|
| **Neural rehabilitation** | Inducing plasticity in paretic finger movement [97] |
| | 3D cursor movement using Tetraplegia ECoG cortical signatures [168] |
| | Stimulating muscle by BCI-driven functional electrical stimulation while bypassing the central nervous system [28] |
| | Treating major depressive disorder patients [105] |
| **Brain-to-brain interface** | Transferring cognitive information regarding words [24], motor [25] and visual [150] stimuli between two human brains directly |
| **Robotics and assistive technologies** | First BCI-controlled robot [143] |
| | Vision guided prosthetic arm for Tetraplegia patients [110] |
| | Controlling humanoid robots [59] |
| | Driving wheelchairs [107], [108] |
| **Space applications** | Monitoring and augmenting astronauts' working capacity [143] |
| | Rehabilitating astronauts by BCI-driven exoskeleton [144] |
| | Enhancing astronauts' functionality, efficiency and safety [145] |
| **Clinical applications** | Treating amyotrophic lateral sclerosis, cerebral palsy, brainstem stroke, spinal cord injuries, muscular dystrophies, or chronic peripheral neuropathies [106] |
| **Miscellaneous** | Controlling virtual reality, video games [67], brain fingerprinting [5], mood assessment [146] and brain painting [147] |